\documentclass[aps,prl,twocolumn,superscriptaddress]{revtex4-1}

\usepackage[LGR,T1]{fontenc}
\usepackage[greek,french]{babel}
\usepackage{epstopdf}
\usepackage{eurosym,hyperref}
\usepackage[dvips]{epsfig}
\usepackage{xcolor,soul}
\usepackage{graphics,graphicx}
\newcommand{\beq}{\begin{equation}}
\newcommand{\eeq}{\end{equation}}
\sethlcolor{yellow}
\setulcolor{red}

\begin{document}
\title{Fast accumulation of ions in a dual trap}

\author{M. R. Kamsap, C. Champenois, J. Pedregosa-Gutierrez, M. Houssin, and M. Knoop}

\affiliation{Aix Marseille Universit\'{e}, CNRS, PIIM, UMR 7345, 13397 Marseille, France}
\email[]{martina.knoop@univ-amu.fr}

\date{\today}
\begin{abstract}Transporting charged particles between different traps has become an important feature in high-precision spectroscopy experiments of different types. In many experiments in atomic and molecular physics,  the optical probing of the ions is not carried out at the same location as the creation or state preparation. In our double linear radio-frequency trap, we have implemented a fast protocol allowing to shuttle large ion clouds very efficiently between traps, in times shorter than a millisecond. Moreover, our shuttling protocol is a  one-way process, allowing to add ions to an existing cloud without loss of the already trapped sample. This feature makes  accumulation possible, resulting in the creation of large ion clouds. Experimental results show, that ion clouds of large size are reached with laser-cooling, however, the described mechanism does not rely on any cooling process.
\end{abstract}

\maketitle

Due to their extremely long storage times and their versatility, radio-frequency (rf) traps are a popular tool for many high-resolution experiments, from quantum information to cold chemistry. They allow the  manipulation and interrogation of ion ensembles from a few units to a million particles, with the possibility to optimise trapping geometries and cloud sizes.  Transporting ions has always been an important ingredient in chemistry or mass spectrometry,  where ions are created in  external sources, and have to be brought to and accumulated in a trap \cite{senko97,herfurth03}. This issue is even more important if the production rate of the ions is low, for example in the case of rare or exotic ions. In this case, the accumulation of ions in the trap is another key-element for the optimisation of the signal-to-noise ratio. The transfer of ions is   central in microwave frequency-metrology experiments, in order to use two separated trapping zones for state-preparation and probing of the ions \cite{prestage07}. Shuttling also gains in importance in quantum information where scalable architectures require the possibility to move ions from on site to another \cite{kielpinski02}.

A major issue during transport is the heating of the transferred atoms or molecules, and as a consequence the large majority  of applied protocols uses a cooling mechanism and or a tailored protocol to limit heating and therefore reduce perturbation and loss of the sample. 

We have recently applied a generic transport protocol, which we have adapted from single-ion translation to the shuttling of a large ion cloud in a macroscopic set-up, and we could demonstrate transfer efficiencies up to 100\% \cite{kamsap15a}. The single-ion transport in micro-traps for quantum information processing  has a particular experimental frame: transport distances are very short (a few 100 $\mu$m), and the objective is to keep the ion(s) in the vibrational ground state. By using many compensation electrodes \cite{wright13}, the transport is designed as a translation of the harmonic potential well (see for example \cite{palmero13}).  Our experimental set-up is different: we aim to shuttle   clouds comprising a few thousand ions over a distance of 23~mm, by using a total  of three DC-electrodes. Despite the differences in size and design, we could adapt  protocols tailored for single ions to  the macroscopic set-up \cite{kamsap15a}. For ion clouds we observed an extra effect which plays the role of  a one-way valve allowing ions to be shuttled from a first trap to a second trap, while forbidding the return of ions already in the second trap. In the present letter, we demonstrate how this "asymmetry" in the working conditions can be exploited to accumulate ions in an ion trap in order to create very large ion clouds. The described process can be optimised by choosing an appropriate set of parameters for the transport between two traps. The physical process involved does not require any cooling mechanism  during or after the shuttling process to assure no loss from the ions already accumulated in the trap. It can then be used to build mixed species cloud, and finds interest in applications in physical chemistry, frequency metrology and antimatter trapping.

\section{One-way transport of ion clouds}\label{sec:trans}
\begin{figure}
\includegraphics[width=85mm]{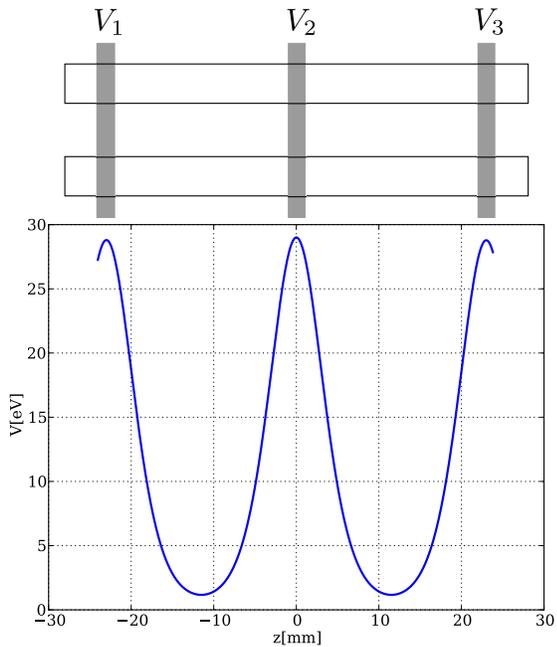}
\caption{Sketch of the ion trap set-up (upper part) and resulting DC potentials (lower part)}
\label{fig:setup}
\end{figure}

The described experiments are carried out in a double linear rf trap being composed of two linear quadrupole traps of radius $r_0 = 3.9$~mm and  length $l = 21$~mm aligned along a common $z$-axis and sharing the same rf electrodes, non segmented. The ion traps are separated by a 2~mm-wide central electrode , where a DC potential barrier is applied. Identical electrodes also close both traps at each end \cite{champenois13bis}, as shown in figure \ref{fig:setup}. In order to transfer ions from the first trap to the second trap, the axial potential  minimum of the trap follows a time varying  transport function, $z_0(t)$, which translates into a temporal variation of the voltage on the central electrode, $V_2(t)$. It has been shown earlier that the analytic form of this function is a key issue in transport \cite{reichle06,bowler12,walther12}. Details of the choice of this  function for the transport of larger clouds are described in \cite{pedregosa15}.

Various applications have different experimental constraints. In our set-up, ensembles of atomic ions (Ca$^+$), are created by photoionization, and laser-cooled. The fluorescence of the photons scattered in the laser-cooling process is recorded by a photomultiplier (PM) in photon-counting mode and an intensified CCD. The detection module is mounted on a slide, and ion clouds can be monitored in every trap by a translation of this slide. Collection efficiencies are identical, and the magnification of the optical set-up changes from 13.2 to 12.9 between both traps. 
Monitoring of the ion number is a central element for the evaluation of the transport efficiencies. In the case of larger ion ensembles, the photon-counting signal from a PM is not a reliable information as it will vary as a function of the cloud's temperature for fixed laser frequencies. In order to be able at any time to enumerate the number of particles in the cloud with precision, we have developed a protocol guaranteeing higher accuracy and reproducibility.

An ion cloud containing more than few hundred atoms can be described by the model of the cold charged fluid, developed for non-neutral plasmas \cite{prasad79}. Then, if the ion cloud is cold and dense enough, like in the liquid phase, the Debye length is small compared to the size of the cloud and its density can be considered as uniform over the whole sample \cite{dubin99, champenois09}. We use this property, and the demonstrated relation between the density and the trapping parameters, to infer ion numbers from the cloud image size. Other characteristics can be deduced from this model, like the aspect ratio of  the ion clouds \cite{turner87}, which has been experimentally confirmed by Hornekaer et al. \cite{hornekaer02}. 

The cloud size can be determined with precision from the CCD images, when the  ions are laser-cooled to low temperatures undergoing a  transition from a thermal gas to a correlated (liquid) state. The recorded images then show an ellipse with a sharp contour edge. An automated fit procedure allows to measure both axes of the observed ellipse with high precision, resulting in an error bar of the relative ion number (comparison between two clouds in the same trapping conditions) lower than 2\% but reaching 5\% for the estimation of the absolute number. 
 The procedure is fast, and as it relies on a fitting procedure of the contour of the cloud, it can still be applied when one dimension of the ion cloud is larger than the observation zone. With the chosen optical  magnification value in our set-up,  we can quantify cloud sizes up to a few 10$^5$ with precision.

\begin{figure}
\includegraphics[width=85mm]{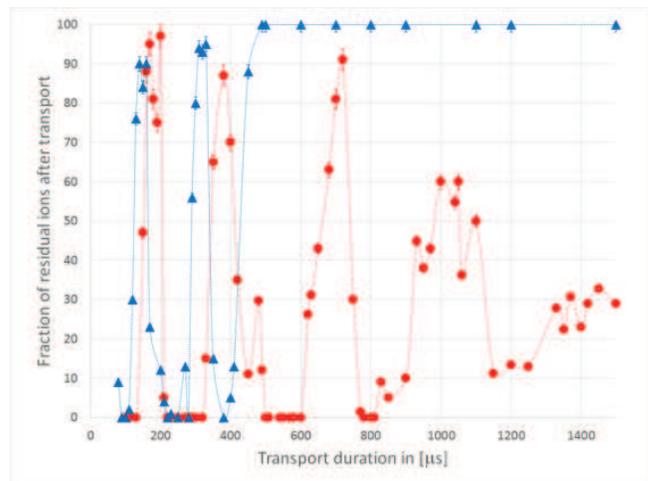}
\caption{Fraction of residual ions in the initial trap when the destination trap is empty, as a function of the duration of the  transport function and of the initial trap :  trap I (red) or trap II (blue). Lines are shown to guide the eye.}
\label{fig:oscillations}
\end{figure}

Cooling to the correlated phase is done before and after every transport, during the transport the cloud can be in the gas or liquid state.  Throughout the entire experiment, the applied transport function is a  variation of hyperbolic-tangent shape \cite{pedregosa15}. For bandpass reasons, the shortest variation applied to the DC-electrodes is 80~$\mu$s. Figure \ref{fig:oscillations} shows in red the fraction of ions still present in trap I after transport, as a function of the duration of the transport function applied to the central DC-electrode. On the same graph, the blue curve shows the results obtained under identical conditions in trap II. Both curves oscillate between 0 and 1, but depending on which trap is used as a starting point, the values of maximum transport differ. 

The large amplitude variations of the curve  in Figure \ref{fig:oscillations} can be explained by the dynamics of the ion cloud. Actually, for a transport duration of 100 $\mu$s starting from trap I, the complete cloud will leave the trap for trap II. At 180 $\mu$s, even though the transport protocol has been applied, 100~\% of the cloud is found at its original location. We have checked experimentally (by opening trap II during the transport) and with numerical simulations, that this corresponds to a situation where  the cloud has left trap I during the transport but was reflected on the first millimetres of trap II, and came back to trap I. This behaviour can be reproduced numerically by small deviations  of the trapping field from the ideal case, and we assume that this is the experimental cause. For longer transport times (> 1~ms), only parts of the cloud come back to trap I, as the spatial spreading of the cloud starts to play a role. On the contrary, for long transport durations, all ions are found in trap II when starting from trap II. This difference for long transport durations as well as the fact that the red and blue curves in Figure \ref{fig:oscillations} do not exactly overlap, is imputed to an asymmetry between the two traps, which is caused by ion creation. Indeed, both  traps have identical size and trapping parameters, but the slight calcium deposit  induced by the calcium beam in trap I, is responsible for a contact potential estimated to 40~mV between opposite rods. In our numerical simulations, we could show that such asymmetry in transfer efficiency can be introduced in ideal traps by adding a local small voltage to one of the rods of one of the traps. 
Experimentally, for a given rf trapping voltage, the described oscillations in transfer probability can be displaced on the axis of transport duration by the variation of the trapping DC-potential (see \cite{kamsap15a} for  details).

\section{Accumulation of ions }

We have used the non-coinciding oscillations of the transfer probability illustrated in Figure \ref{fig:oscillations} to proceed to a "no-return" transport protocol which allows to add ions from trap I to trap II, without loosing the cloud already trapped in trap II.  By choosing a transport duration for which the  transport protocol results in a very different result depending on the initial trap, we can create a situation where ions are accumulated in one trapping zone. For the conditions of Figure \ref{fig:oscillations} this corresponds for example for durations 300, 550, and 780~$\mu$s. All ions will then leave trap I, and no ion will leave trap II. 

\begin{figure}
\includegraphics[width=85mm]{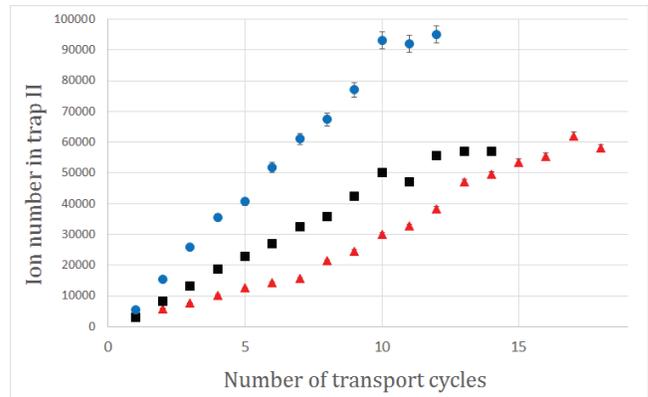}
\caption{Number of ions in trap II as a function of the  number of transport cycles. The accumulated cloud in trap II is in the gas phase (blue dots and red diamonds) or in the liquid phase (black squares). The transported clouds have different size : creation time of 25 s (blue dots) or 15 s (red diamonds and black squares)}
\label{fig:accum}
\end{figure}

The resulting accumulation of ions is illustrated in Figure \ref{fig:accum} for various transported cloud sizes and accumulated cloud states. The graph reports the ion number in trap II as a function of the transport cycle  number for a fixed transport duration. We call a transport cycle the creation in trap I followed by the transport to trap II. Depending on the experimental parameters, we have created ion clouds of a few thousand ions in trap I. The reproducibility of the creation for a given set of conditions is better then 4 \%. The cloud is transferred to trap II, then measured with the protocol described above.  After this, a new cloud is  created in trap I for the next transport cycle. Figure  \ref{fig:accum} shows how the cloud in trap II builds up as a function of the number of transport cycles. We have tested if the state of the initial cloud in trap I has an influence on the accumulation feature. Figure \ref{fig:accum} illustrates accumulation for different starting configurations: just before transport the  cloud in trap I can be in the gas phase or in the liquid phase. The clouds in trap I and trap II see the same cooling laser beams and trapping potential, therefore we can assume that the cloud in trap I  is in the same state as the accumulated cloud in trap II (if the difference in size is not too important). 

The growth of the ion cloud in trap II can be described as being linear  for at least 10 cycles.  We then observe a saturation of the size which depends on the cooling laser power and the RF and DC trapping potential. Higher laser-power and a lower trapping potential allow to reach larger clouds, compatible with a limitation by the temperature of the accumulated cloud.  We apply collimated cooling laser beams at 397~nm of a typical diameter of 2.5~mm with a power of 3~mW and varied the trapping conditions up to a Mathieu parameter of $q_x$=0.10.  
 
This experiment describes a genuine accumulation process, meaning that ions can be added to an already existing cloud in a trap. This is an important fact for all experiments where ions are rare or difficult to produce. It allows to grow ion clouds in a pulsed regime by adding particles to an existing cloud. To our knowledge, the large majority of existing experiments uses the term accumulation in the sense of a variable integration time during initial creation or loading of a trap, which describes a different creation process.

\section{Accumulation without cooling}
\begin{figure}
\includegraphics[width=85mm]{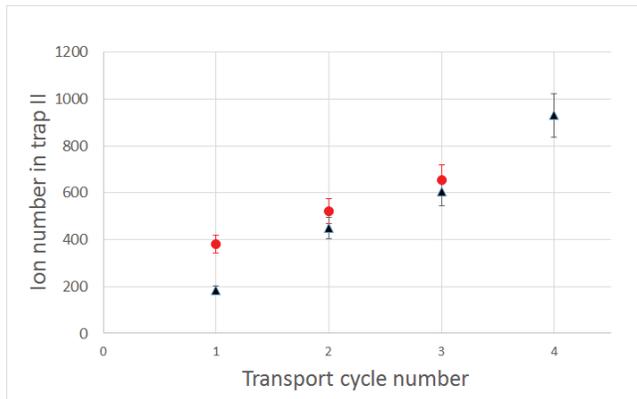}
\caption{Transport of ions without any cooling process. The graph shows the cumulated ion number in trap II as a function of the transfer cycle. Small ion clouds are used, approximately 900 ions (black triangles) and 1300 ions (red dots). }
\label{fig:accum_woc}
\end{figure}

Being able to operate without a cooling process is an advantage for many experiments. We have realised a sequence of transport for small ion clouds without laser-cooling. The reported experiments are carried out under ultra-high vacuum conditions at pressures below 3$\cdot 10^{-9}$, in the absence of buffer-gas cooling (or any other cooling mechanism).  Results are plotted on Figure \ref{fig:accum_woc}. As before, they show the cumulated ion number in trap II as a function of the number of transfer cycles, even if the cloud sizes are smaller than in the laser-cooled case. For these small clouds we also evidence a linear increase in ion number in trap II, showing that the mechanism of transport and accumulation is efficient even without the application of a cooling process. This is an important step for transporting charged particles, as the vast majority of experiments relies on buffer-gas or laser-cooling during or after the transport in order to damp the transport-induced heating.

\section{Outlook}
We have demonstrated accumulation of ions in a dual linear trap, with and without laser cooling. The method relies on an asymmetry between the two traps, responsible for a different timing for one way and return transfer efficiency. Accumulation of ions in a trap is a key-element in many experiments working with  ion clouds of different size, in particular where the rate of ion creation is low. The described mechanism can  also be  adapted to a multi-species scenario allowing to load different types of charged atoms or molecules in the same trap. This kind of ensemble gains importance in interaction studies, it is in particularly employed in sympathetic cooling and quantum logic protocols \cite{schmidt05}.

\acknowledgments {This research has been carried out under contract n$^{\circ}$116279 with the French spatial agency (CNES) and ANR-08-JCJC-0053-01 from Agence Nationale de la Recherche; MRK  acknowledges financial support from CNES and R\'{e}gion Provence-Alpes-C\^{o}te d'Azur.}

\bibliographystyle{apsrev4-1}

%

\end{document}